# Development of mirrors made of chemically tempered glass foils for future X-ray telescopes

**Bianca Salmaso · Marta Civitani · Claudia Brizzolari · Stefano Basso · Mauro Ghigo · Giovanni Pareschi · Daniele Spiga · Laura Proserpio · Yves Suppiger**

**Abstract** Thin slumped glass foils are considered good candidates for the realization of future X-ray telescopes with large effective area and high spatial resolution. However, the hot slumping process affects the glass strength, and this can be an issue during the launch of the satellite because of the high kinematical and static loads occurring during that phase. In the present work we have investigated the possible use of Gorilla® glass (produced by Corning®), a chemical tempered glass that, thanks to its strength characteristics, would be ideal. The un-tempered glass foils were curved by means of an innovative hot slumping technique and subsequently chemically tempered. In this paper we show that the chemical tempering process applied to Gorilla® glass foils does not affect the surface micro-roughness of the mirrors. On the other end, the stress introduced by the tempering process causes a reduction in the amplitude of the longitudinal profile errors with a lateral size close to the mirror length. The effect of the overall shape changes in the final resolution performance of the glass mirrors was studied by simulating the glass foils integration with our innovative approach based on glass reinforcing ribs. The preliminary tests performed so far suggest that this approach has the potential to be applied to the X-ray telescopes of the next generation.

**Keywords** X-ray optics, X-ray mirrors, Glass, Glass strength, Slumping

## 1 Introduction

Glass foils thinner than 1 mm are presently considered a very valid solution for making the large X-ray telescopes of the future, that will have to couple a good angular resolution (a few arcsec, Half Energy Width, HEW) to a high throughput. X-ray astronomical mirrors based on thin glass foils were firstly introduced for the HEFT balloon experiments for small size hard X-ray optics (Koglin 2004), and then used in the NuSTAR mission (Craig 2011). Concerning large size X-ray optics, they were firstly considered by NASA during the development of the Constellation-X optics (Zhang 2007) and then studied again by NASA for the IXO project (Zhang 2010). In Europe, at first in the context of XEUS (Fredrich 2006; Ghigo 2008) and afterwards in the context of IXO (Pareschi 2011; Winter 2010) the thin glass approach was also developed, with very promising results. Now the Athena mission (Nandra 2014) has been approved for the L2 slot in the Cosmic Vision program. At this regard, in addition to the baseline solution based on Silicon Pore Optics (SPO), thin glass foils are also considered a viable alternative. Finally, thin glass foils with piezo-electric actuators are being developed by a group at SAO/CfA in USA to manufacture the mirrors of the SMART-X mission (Reid 2014) and are also studied in Italy (Spiga 2014).

In our laboratories we have introduced a novel approach for the production of the X-ray telescopes. It foresees the use of thin (0.4 mm) glass foils curved by hot slumping process in a cylindrical shape and then assembled into the final Wolter-I configuration, a typical configuration for the X-ray telescopes including a paraboloid and an hyperboloid (Van Speybroeck & Chase 1972), onto which X-rays make two reflections in sequence. This research project was supported until 2013 by the European Space Agency as a backup development of the IXO optics; it was then continued in our group for the study of the Athena optics. Four prototypes have been realized so far and tested in X-rays at PANTER/MPE (Burwitz 2013), showing an angular resolution improvement from the 80 arcsec of the first prototype to 22 arcsec, as of today the best result obtained for the entire module in full illumination (Civitani et al. 2014).

Our hot slumping process is assisted by pressure, in order to help the glass to replicate the mould shape (Proserpio 2011). It is a direct replication approach, which means that the surface in contact with the mould is the one that will be used for the reflection. As the thermal cycle and the contact of the glass with the mould might affect the strength of the glasses, our group carried out an intense study to check the effect of the slumping process on the strength of the glasses (Proserpio 2013, 2014). This work was committed by the European Space Agency, to be addressed with a statistical approach (Weibull statistics) due to the large number of IXO glass mirror segments (16.560). Two reference values for the reliability of the whole mirror assembly were provided by ESA, the stricter being 99.99% and the looser 99.00%. These numbers represent the probability to avoid any catastrophic event. A breakage is defined to be catastrophic when causing the detachment of a large number of glass fragments, which could float in the spacecraft leading to the complete mission failure. Due to the large number of the IXO mirror foils, and considered that the usual glass strength distributions show a large scattering behavior, the evaluation of the Weibull parameters characterizing the full mirror foil's population required a very large number of specimens. At this scope, more than 200 samples of Schott D263 glass were used to perform double-ring destructive tests. . As a main result of the study, it has been found that the minimum tensile stress at breaking point, measured by destructive double-ring tests, was 82 MPa for curved glasses, just slightly smaller if compared to the 117 MPa for the as-delivered flat glasses before slumping. The reduction of the glass strength by the slumping process could ensure a survival probability of 99.00% for the IXO telescope, while the most severe survival level of 99.99% could not be reached.

These numbers appears very severe, specially when compared with the NuSTAR telescope, that is also based on optics made with slumped glass foils connected by ribs. The NuSTAR flight optics successfully survived vibration tests (Craig 2011). Moreover, more than 400 representative witness coupons were produced to test the NuSTAR manufacturing process: the bondings, not the glass foils, were defined as the primary mode of failure. Nevertheless we have to consider that the number of the IXO glass foils is much larger (16.560) than the one for NuSTAR (2.376 foils for each of the two optics) increasing the probability than one single foil would break. Moreover the NuSTAR entrance and exit apertures of the optics modules are protected with thermal covers (Craig 2011), that would also prevent possible glass fragments to float in the spacecraft in case of breaks. The comparison of the loads that the two optics had/should sustain is not trivial. Equivalent static loads at launch of ±70g in the longitudinal and ±55g in the later direction where considered for IXO to compute the survival probability (Parodi 2011): these values were obtained rounding the algebraic sum of the quasi-static transient loads (longitudinal = 18g, lateral = 4g), provided by ESA, and random vibrations effects managed as equivalent static loads (longitudinal = ±50g, lateral = ±50g). No equivalent numbers are reported in the literature for NuSTAR, to our knowledge. Anyway, being the telescope structures totally different, the loads on the telescope would be different.

Relying on the numbers obtained in our research (99.00% survival probability of the whole mirror module, to be compared with the most stringent 99.99%) we have investigated possible solutions to increase the survival probability of the mirror assembly. A natural choice is to re-enhance the slumped glass strength with a step of chemical tempering after slumping. The chemical tempering, in fact, changes the surface properties of the glass introducing a compressive layer that enormously strengthens the glass. As the increase in strength is strongly dependent on the glass type, we have selected for our research the Corning® Gorilla® glass, well known for its capability to develop high compressive strength. The compressive strength after chemical tempering for the Gorilla® glass is reported to be higher than 800 MPa (Corning spec®, Gorilla 2). For this research, we used the Gorilla® glass 2, with thickness 0.55 mm, the thinnest configuration available when we performed the study. The strength of the un-tempered and tempered Gorilla® glass 2 were compared at Corning with double-ring destructive tests on flat glass foils of 1 mm thickness (Corning private communication). The measured failure loads, at 62.5% failure probability, were founds to be 2.746 and 3.687 N for the un-tempered and tempered glass respectively, showing the net increase in strength after the chemical tempering. No tests were done at Corning on curved samples, despite the fact that curved Gorilla® glass are commercially used. We did not yet measure the strength of our slumped and tempered Gorilla® samples, to compare with the slumped and un-tempered samples. This could be a subject for a following paper.

The key question for us to be addressed was the effect of the chemical tempering on glass roughness and shape, to define the possibility to use the Gorilla® glass for X-ray optics. Therefore we have carried out tests to define the slumping process by using a cylindrical mould in Zerodur K20, with a curvature radius of 1 m; the glass were cut by $CO_2$ laser at MDI-Schott, to ensure the survival during the tempering process; the chemical tempering was performed at EuropTec in Oftringen, a company skilled at the production of glasses of different geometries and surface properties; all the metrological characterizations were performed in our laboratories.

This paper is organized as following. In section 2 we describe our Slumped Glass Optics (SGO) technology. In section 3 the chemical strengthening process is described and the comparison of the thermal properties of the different glass types used in our laboratory are presented. In section 4 we describe the metrology instrumentation used. In section 5 we deal with a critical analysis of the roughness achievable on tempered Gorilla® glass foils. In section 6 we describe the slumping process specifically developed for obtaining X-ray mirrors based on the Gorilla® glass. In section 7 we present the cutting and tempering process of our sample glasses. In section 8 we show the roughness data after the process of slumping and tempering. In section 9 we compare the shape measurements of the slumped foils before and after tempering; we also present the expected image resolution of the segments, achievable with our integration setup, showing that the tempering process does not introduce a significant degradation. Finally, in section 10 we summarize all the results of the performed work and discuss about possible future developments.

## 2 The SGO: hot slumping assisted by pressure and integration with IMA (Integration Machine)

The optical module of the Athena X-ray telescope will have a diameter of about 3 m, therefore its production has to be based on a modular approach. Using the SGO technique, thin glasses are stacked into an X-ray Optical Unit (XOU) and then assembled as sectors in the final mirror assembly (Fig. 1) (Basso 2014).

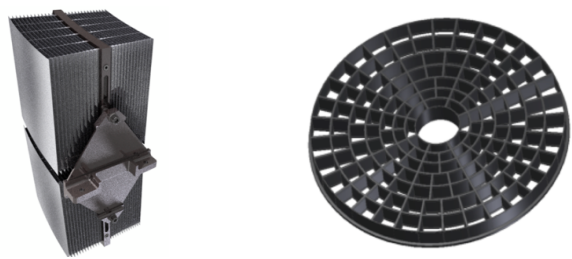

**Fig. 1** Left: the X-ray Optical Unit. Right: the final assembly with the sectors for the XOUs.

Each individual glass foil has to be curved in a furnace with a proper thermal cycle that enables a good surface roughness and a proper curvature. In Fig. 2 the furnace, the slumping mould, the glass and the muffle used to enable the pressure application are shown. The glass is slumped with larger dimension than the mould, and also acts as a membrane to divide the muffle chamber in two parts: the pressure is exerted by vacuum suction from the lower part. This setup has been patented (TO2013A000687, 12/08/2013). The slumped glass has then to be trimmed to the final dimension, 200 x 200 mm$^2$. The thickness of the glass used so far is 0.4 mm.

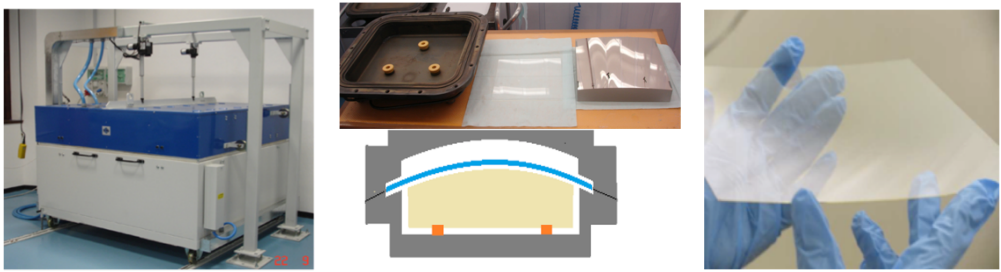

**Fig. 2** Left: the furnace at our laboratories. Center: the muffle, the glass, the slumping mould and the slumping set up. Right: the slumped glass after the trimming.

The glass foils, after slumping, assume a cylindrical shape and have to be integrated into the supporting structure. During the integration process we impart to the foil the final Wolter I configuration. This is performed via an ad-hoc designed and developed machine named Integration MAchine (IMA) (Fig.3, Civitani 2011). The glass foils are stacked and glued with glass spacers (called ribs) that endow the structure with more stiffness and freeze into the glasses the Wolter-I profiles of the integration mould, thus correcting for residual low-frequency error profiles (Civitani 2013). Finite Element Analysis was performed with the ANSYS software (Parodi 2011) to evaluate the capability of the process to correct initial errors with different spatial wavelengths. Long-scale errors can be corrected almost entirely, whereas errors over a few cm-scale remain almost unchanged after the integration. The correction capability of our integration system was previously proven by X-ray tests on our prototypes, integrated on 200 × 200 mm$^2$ size with 6 ribs (Civitani et al. 2014). The best result obtained so far for the entire module from X-ray calibrations in full illumination is 22 arcsec (Civitani et al. 2014) with Schott AF32 glass foils. This result was limited by profiles errors in the slumped glass foils and (even if at a lower level) by the integration moulds that have been fabricated with a not yet perfect reference surface. While we are working to improve these aspects, our simulations prior to perform the integration of tiles and reverse-engineering analysis after the X-ray tests tell us that the ribs mounting scheme properly correct the low frequency errors and doesn't introduce any important other error in the HEW of the tiles.

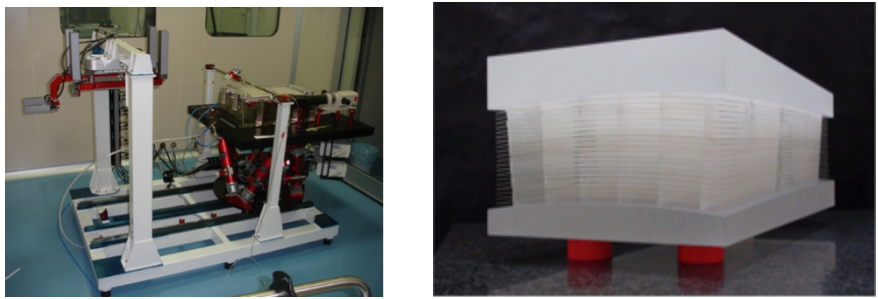

**Fig. 3** Left: the Integration MAchine. Right: a complete glass stack with the reinforcing ribs.

## 3 Chemical strengthening

As the glass is a brittle material and the telescope have to withstand the intense stresses experienced during the launch, the strength of the glasses must be carefully considered. If the glass surface were flawless, the strength would approach the theoretical value of 20 GPa (Holloway 1973), as computed from the bonding energies of silicon and oxygen. Indeed, the glass surfaces are subject to damage during manufacturing, leading to increased stress near the flaws according to Eq. 1 (Inglis 1913, Griffith 1924; Kulp 2012):

$$\sigma = 2\sigma_A \left(\frac{l}{r}\right)^{\frac{1}{2}} \qquad (1)$$

being $\sigma_A$ the applied stress, $\sigma$ the stress on the flaw tip, $l$ the flaw length, $r$ the flaw tip radius, assumed to be equal to the interatomic distance (~ 1 Å). When the applied stress causes $\sigma$ to exceed the theoretical strength, a failure will occur. Typical defects occurring on the glasses have $l$ ranging from 1 to 100 μm, returning maximum applicable stresses from 100 to 10 MPa, respectively.

A further slight damage is introduced by the hot slumping process, contributing to a reduction of the glass foil resistance, as already demonstrated with D263 glass foils (Proserpio 2014). Therefore, we considered the possibility to re-enhance the glass strength after the slumping.

There are several techniques to improve the glass strength, one of which is chemical strengthening. In this method the glass is dipped in a molten salt bath (Uhlmann 1980), with temperature ranging from 350 to 550 °C. During the process, at the interface between the glass and the molten salt, two types of alkali ions exchange places with each other (Adamson 1997), for example a sodium ion from the glass will exchange places with a potassium ion from the salt. The ion exchange is temperature and time dependent. As the potassium ion occupies a larger volume than sodium, a compressive stress will be developed in the part of the glass where the ion exchange process occurred. One of the most important parameters is the depth to which this process occurs, because the strengthening process effectiveness increases with the depth. In addition to time and temperature, the composition of the glass and the ion exchange species play an important role in the ion exchange process. It has been shown (Uhlmann 1980), that alkali-alumina-silicate glasses (like the Gorilla® glass) are the most receptive to the ion exchange process. Borosilicate glasses (like D263 glass) instead have been shown to be a less desirable composition. Although the process can still be carried out using this composition, it occurs at a slower rate, and yields lower strengths and exchange depths.

The Gorilla® glass was selected for this investigation among three available aluminosilicate thin glasses (Corning® Gorilla® glass, SCHOTT Xensation™ Cover glass and AGC Dragontrail™) because of the particular Corning® forming process (fusion for the Gorilla® glass, microfloat for the SCHOTT Xensation™ and float for the AGC Dragontrail™), which, in our experience, gives the best results in terms of microroughness (Salmaso 2014).

The Gorilla® glass generation 2 was used because it exhibits the same resistance characteristics than the generation 1, even if it is thinner by 25 %. The used thickness was 0.55 mm, the thinnest at the time of the research. It is reported (Corning spec®, Gorilla 2), that this glass, after tempering, develops a compressive stress beyond 800 MPa with a Depth Of Layer (DOF) larger than 40 μm, to be compared to typical values of minimum tensile stress at breaking point ranging from 100 to 150 MPa for non-tempered glasses (Kulp 2012).

In Table 1 we report a comparison of some key parameters of the Gorilla® glass with other glass kinds used in our laboratories for the slumping technology development for X-ray telescopes. Schott D263 glass was used in the first part of our study and the strength of the slumped glasses was fully characterized (Proserpio 2013, 2014); it was also used for the NuSTAR optics (Craig 2011). Schott AF32 glass was then adopted in our study as it has provided better results owing to the better Coefficient of Thermal Expansion (CTE) matching with the Zerodur K20 mould in use (Civitani 2013); unfortunately, AF32 glass is not suitable for chemical tempering as it is a alkali-free glass. Finally, Corning Eagle glass was adopted because it has the same CTE of the AF32 glass, but a better quality surface in terms of waviness and thickness uniformity (Salmaso 2014).

**Table 1** Parameter comparison for some glass brands

|  | Gorilla-2[1] | D263[2] | AF32[3] | Eagle XG[4] |
|---|---|---|---|---|
| Producer | Corning | SCHOTT | SCHOTT | Corning |
| Type of glass | Aluminosilicate | Borosilicate | Aluminoborosilicate Alkali free | Alkaline Earth Boro Aluminosilicate |
| n @ 589 nm ($n_D$) | 1.505 @ 590nm | 1.523 | 1.51 | 1.51 |
| Softening point | 852 | 736 | 969 | 971 |
| Annealing point | 613 | 557 | 728 | 722 |
| Strain point | 563 | 529 | 686 | 669 |
| Slumping temperature @ INAF-OAB | 620 | 570 | 750 | 750 |
| Thickness [mm] | 0.55 | 0.4 | 0.4 | 0.4 |
| CTE ($10^{-6}$K) | 8.1 (th=0.55mm) | 7.2 | 3.2 | 3.17 |
| Chemical tempering | Possible | Possible | Not possible | Not reported |

[1] http://cgg.dev-box.org/uploads/kcfinder/files/Gorilla%20LCG%20PI%20Sheet_GG1_02112013.pdf
[2] http://www.howardglass.com/pdf/d263t_eco.pdf
[3] http://www.howardglass.com/pdf/af_32.pdf
[4] http://www.howardglass.com/pdf/eagle_xg.pdf

## 4 The metrology apparatus

To measure the 1D profiles of the slumped glasses, we used the CHRocodile 3300 sensor head (by Precitec), mounted on the Long Trace Profilometer (LTP), operated at the INAF/OAB laboratories. To subtract the unwanted LTP stage movement in the vertical direction, a calibrated rod was also scanned as reference during the measurements with the Rodenstock RM600-S sensor (Fig. 4). The CHRocodile sensor was used in thickness mode to simultaneously record the profile of both sides of the foil. The samples, supported on three bearing points, were measured on few longitudinal scans of the cylinder with the optical side downwards and upwards:, the two measurements of the optical side of the glass were subtracted in order to remove gravity and the bearing point deformations..

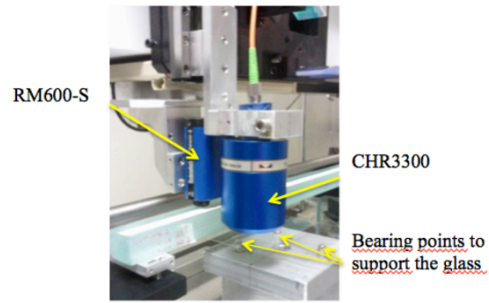

**Fig. 4** Setup for 1D profiles measurements: the CHRocodile 3300 sensor head mounted on the LTP and the RM600-S sensor measuring the calibrated rod to subtract the LTP stage errors. The glass is supported horizontally on three points.

To measure the 2D maps, the Characterization Universal Profilometer (CUP) (Civitani 2010), developed in our laboratories, was used. This instrument measures the distance of the surface under test from a rigid reference plane (Fig. 5). The CHRocodile 660 sensor head (20 nm resolution), mounted on the xyz translational stage measures the distance from the optical surface and provides a feedback to the precision stage motors in order to keep the sensor distance constant ("null sensor") within the sensor resolution; at the same time, a SIOS triple axis interferometer measures the distance from a calibrated reference mirror. The samples were vertically supported on three points, a configuration that minimizes the effect of gravity deformation. The measurements are carried out in a thermally controlled clean room (±0.2°C) to minimize thermal expansion variations of the carriages that might introduce a spurious component in the measured profile.

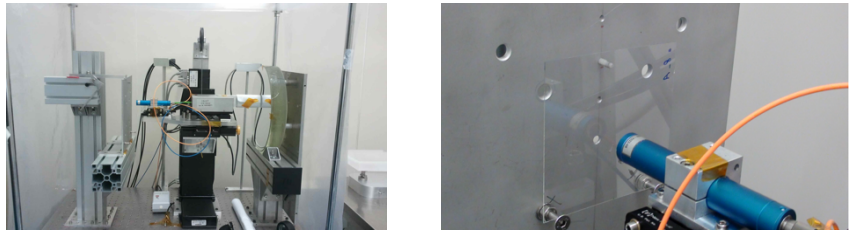

**Fig. 5** Setup for the 2D profiles measurements. Left: the CUP setup. Right: the CHRocodile 660 sensor head mounted on the CUP. The glass is supported vertically on three points.

The roughness of the glass surface, at spatial wavelengths ranging from 2.5 mm to 10 μm, was measured via white-light interferometry using the WYKO microscope available at INAF/OAB, with a vertical resolution of few angström. On the low frequency side, the roughness characterization overlaps the spectral band covered by the profile measurements. On the high frequency side, a roughness measurement is not essential because – at X-ray energies near 1 keV and incidence angles near 0.7 deg – the high-frequency roughness has a negligible impact on the angular resolution to within large limits of rms values.

## 5 Roughness characterization of flat tempered Gorilla® glass

In order to ascertain if the surface roughness of the tempered Gorilla® glasses does not degrade the high angular resolution required to slumped glass mirrors, we measured a flat tempered Gorilla® glass with the WYKO interferometer. The Power Spectral Density (PSD), computed from these measurements, was compared with the ones of the brand new glasses that we are evaluating for the production of X-ray optics, namely D263, AF32 and Eagle XG glasses (Fig. 6). All measured PSDs are compared with two requirements, shown as dotted lines. Our old PSD requirement (dotted black line in Fig.6) is derived from the demand that the HEW degradation caused by X-ray scattering at 1 keV and 0.7 deg is below 1 arcsec. The PSD requirement function was then computed using an analytical model developed by Spiga (Spiga 2007), able to predict the contribution of the X-ray scattering to the HEW of a grazing incidence X-ray optics. Using this formalism one is able to derive the PSD that a surface should have in order to satisfy a given HEW. Our new requirements in terms of PSD behavior is also over-plotted in Fig.6 (dotted pink line). This represents the PSD of the brand new glass foils, compliant with a HEW degradation due to scattering < 1 arcsec up to 6 keV (as e.g. requested in the Athena case) and not just at 1 keV as it was for our early study. The measurements show that the chemical tempering does not introduce any degradation on the glass surface micro-roughness with respect to non-tempered glass foils.

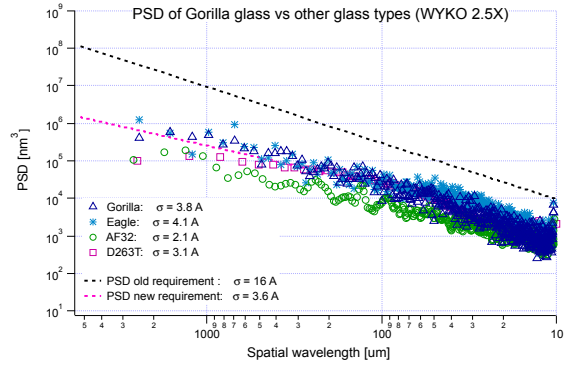

**Fig. 6** PSD comparison of the tempered Gorilla® glass with Eagle, AF32 and D263 non-tempered glasses. All the measurements were taken before slumping.

**6 Slumping of the Gorilla® glass**

The slumping technology consists of replicating the shape of a mould on a thin glass foil, by placing the glass foil onto the mould inside an oven, and bringing the glass to a temperature high enough to soften and slump over the mould. The temperature to be reached depends on the characteristics of the glass being slumped, namely its softening, annealing and strain points that are defined as the temperatures at which the glass viscosity takes on characteristic values. At high temperature, the glass deforms and changes shape until it comes into contact with the mould. In our laboratories, we have developed a direct slumping approach, in which the glass surface that goes into contact with the mould is the one that will be used for the reflection. Moreover, the mould-foil close touch is forced by pressure application (Proserpio 2011). The mould and the glass foil are closed in a muffle, made of AISI 310 stainless steel. The glass foil has a wider size than the final one: this is necessary to create two chambers in the muffle and exert a differential pressure via air suction in the section that contains the mould. The pressure control is achieved through a controller from MKS Instruments Inc, used with a precision capacitance manometer from the same manufacturer. The muffle is positioned inside the slumping oven (by Teknokilns) equipped with five different heating zones in order to minimize temperature gradients throughout the mould. The thermal cycle and the applied pressure are tuned for a good shape replication of the mould and a minimal roughness transferred from the mould to the glass foil. At the end of the thermal cycle the glass foils have to be trimmed to the desired final dimension.

The mould used for the slumping of the Gorilla® glass foils was the same used for the production of the prototypes made on D263, AF32 and Eagle XG (Civitani et al. 2014): a cylindrical mould, with base dimensions of $250 \times 250$ mm$^2$ and radius of curvature of 1 m – resulting in a of 7.8 mm sagitta. The mould is made of Zerodur K20, a semi-crystalline material that does not stick to the glasses of the adopted types across the applied thermal cycle. This material was selected owing to the good matching of Coefficient of Thermal Expansion (CTE) with the AF32 and Eagle XG glasses ($CTE_{K20} = 2.2 \cdot 10^{-6}$/K, $CTE_{AF32} = 3.2 \cdot 10^{-6}$/K, $CTE_{Eagle} = 3.17 \cdot 10^{-6}$/K), thus assuring a good coupling of the two surfaces, especially during the cooling phase when the shape is frozen into the glass. The CTE matching is actually not ideal for the Gorilla® glass foils ($CTE_{Gorilla} = 8.1 \cdot 10^{-6}$/K), but the mould was anyway used to the sole aim of investigating the effect of chemical tempering. At this regard, we are currently investigating the hot slumping of the Gorilla® glass foils on sample moulds made of Alumina, whose CTE ($CTE_{Al2O3} = 7.5 \cdot 10^{-6}$/K) value matches much better the Gorilla's.

In order to assure the non-sticking behavior of the Zerodur K20 with the Gorilla® glass, preliminary slumping tests were performed on small flat Zerodur K20 sample moulds. Different combinations of temperatures (600 to 650 ºC) and pressure (0 to 50 g/cm$^2$) were tested and the final thermal cycle was defined (Table 2) with a pressure application of 50 g/cm$^2$, starting just after reaching the soaking temperature 620 °C, and ending at temperature below 200 °C in the cooling-down phase. Four Gorilla® glass foils were slumped on the cylindrical Zerodur K20 using the thermal cycle described in Table 2.

**Table 2** Thermal cycle for the slumping of the Gorilla® glass

| Step | Temperature [°C] | Time [h] |
|---|---|---|
| 1 | 25 → 620 | 10 |
| 2 | Hold at 620 | 2 |
| 3 | 620 → 602 | 4 |
| 4 | Hold at 602 | 2 |
| 6 | 602 → 530 | 10 |
| 7 | 530 → 450 | 5 |

Owing to the quartz glass window present on the oven, it was possible to record the evolution of the interference fringes in the gap between the glass foil and the mould during the thermal cycle. The images at significant process times are reported in Fig. 7. They show the contact of the glass to the mould with the application of pressure (Fig. 7.b) and the contact release during the cooling-down phase.

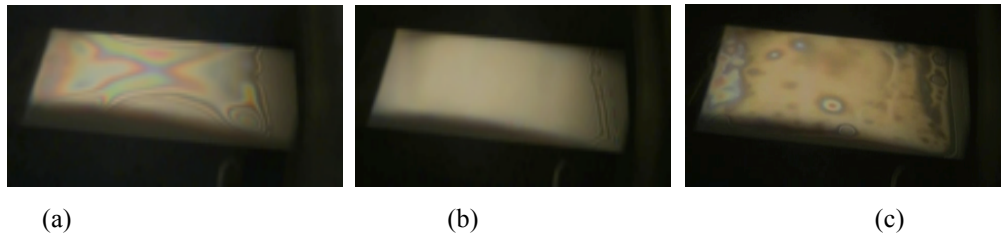

(a) (b) (c)

**Fig. 7** The interference fringes of the glass and the slumping mould recorded during the thermal cycle (a) pressure application begins at $T=620°C$ (b) After 2 h held at $T=620°C$ (c) After 10 h of ramp down, at $T=530°C$.

## 7 Cutting and tempering of the slumped Gorilla® glass

The glass foils, slumped to size of $340 \times 340$ mm$^2$, were preliminarily cut to approx. $230 \times 230$ mm$^2$. The cut was done propagating an initial crack in the glass foil by means of a hot tip. In this configuration, the four slumped foils have been measured with the CHR mounted on the LTP (see Sect. 4) to evaluate the glass profile deviation from the mould, and therefore the foils appropriateness for our slumping process, despite the difference in CTE.

The slumped glass foils were cut by $CO_2$ laser at MDI-Schott in Germany in four $100 \times 100$ mm$^2$ foils (Fig. 8) in order to test the survival of the glass foil upon the chemical tempering process. In fact, the defects introduced by the cutting process reduce the strength of the glass and therefore also the survival probability of the glass foils during the tempering process. The $CO_2$ laser is reported to be the best cutting process to avoid edges micro-cracking and chipping (Coherent Inc. 2008). The chemical tempering of these slumped glass foils was done at EuropTec in their standard production baths. None of the glass foils broke down during the tempering process.

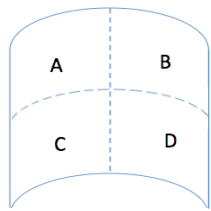

**Fig. 8** The slumped glass foils were cut into four $100 \times 100$ mm$^2$ glasses by $CO_2$ laser.

In the following sections, since we are interested only in checking the chemical tempering effect on good-shaped glasses, especially in the centimeter spectral region of spatial wavelengths, only one glass foil (named G7) is considered. We refer to the four sub-foils obtained from splitting the G7 glass as G7A, G7B, G7C, and G7D.

## 8 Roughness characterization of slumped tempered Gorilla® glass

The surface roughness of the slumped and tempered Gorilla® glass sub-foils, obtained form the G7 foil, were measured with the WYKO interferometer on their central axis, and the results were compared to the surface roughness measurements obtained from the flat tempered Gorilla® glass foil.

The measurements show (Fig. 9) that the entire process of slumping and tempering does not introduce any degradation on the glass surface micro-roughness, and it is therefore suitable for the production of X-ray glass mirrors.

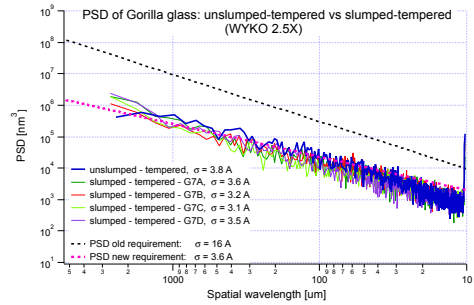

**Fig. 9** PSD of the slumped and tempered Gorilla® glass foils G7A, G7B, G7C and G7D, compared to the PSD of the flat tempered Gorilla® glass foil.

## 9 Shape characterizations

The four samples, obtained from the G7 glass foil, were characterized in shape before and after the chemical tempering, using (Sect. 4) both the CHR - LTP (mono-dimensional profiles) and the CUP (bi-dimensional surface map).

*9.1 1-dimensional profiles measurements: CHR on LTP*

Three profiles were measured with the CHR - LTP before and after chemical tempering: the actual profiles are derived from the difference of the measurements with concavity up and down, in order to subtract the gravity and the bearing deformations. The scans are at the central position (denoted as C), and 30 mm left and right of the foil center (denoted as 30R and 30L).

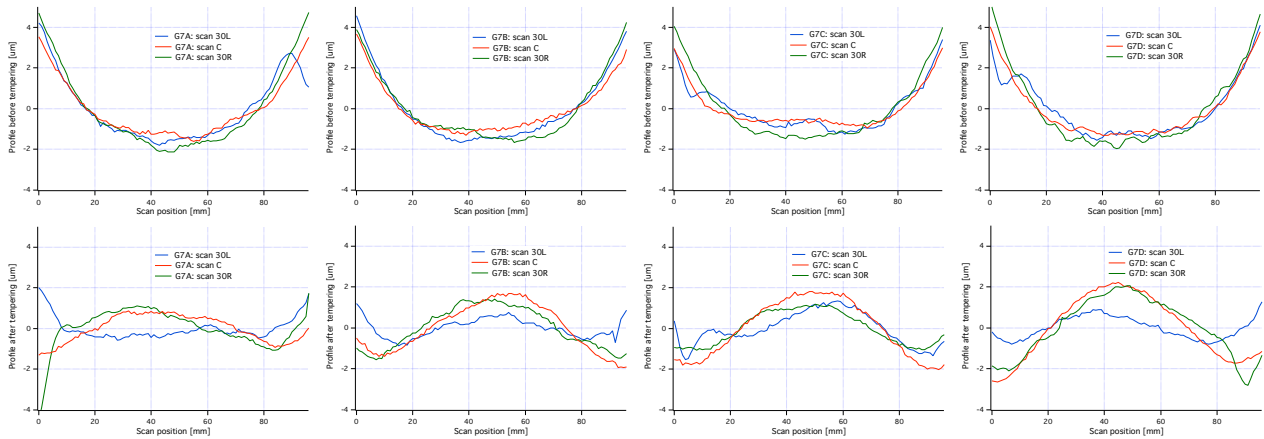

**Fig. 10** Profiles computed from up-down CHR measurements of the slumped Gorilla® glasses. (top) profiles before tempering; (bottom) profiles after tempering. From left to right: the samples A, B, C and D.

Fig. 10 shows that the chemical tempering reverses the concavity of the profile error in the longitudinal direction, with a reduction of the peak-to-valley (PV). Even though the integration procedure acts as a damping of the low-frequency errors (Parodi 2011), the remaining amplitude of the low-frequency error still represents a contribution to the angular resolution degradation. To understand whether the mid-frequency errors are real profile features or metrology artifacts, a new measuring method was developed: the LTP is used in its usual slope-detecting configuration and the reflection of the back surface of the glass is suppressed by painting it with First Contact® (by Photonic Cleaning) (Salmaso 2014). The glass is measured only with concavity upwards and therefore the profile contains also the deformation caused by the gravity and the bearing points, which are removed subtracting the profile simulated by finite element analysis applied to a perfect cylindrical mirror. This method could only be used for the slumped and tempered samples. For the glass foils before slumping, this solution had not been found yet at the measurement time, and the LTP-CHR setup was adopted (Sect. 4). Fig. 11 compares the LTP-CHR and the LTP measure of the G7-D central scan. It proves that the mid-frequency errors detected with the LTP-CHR are spurious.

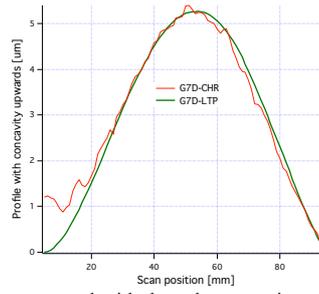

**Fig. 11** Profiles of the G7-D sample after tempering, measured with the sole concavity upwards: in red the CHR scan, in green the LTP scan.

*9.2 2-dimensional profiles measurements: CUP*

The surface profile was also measured with the CUP, with the setup described in Sect. 4. Fig. 12 shows the figure errors, with respect to the nominal cylindrical surface, of the four samples before and after tempering.

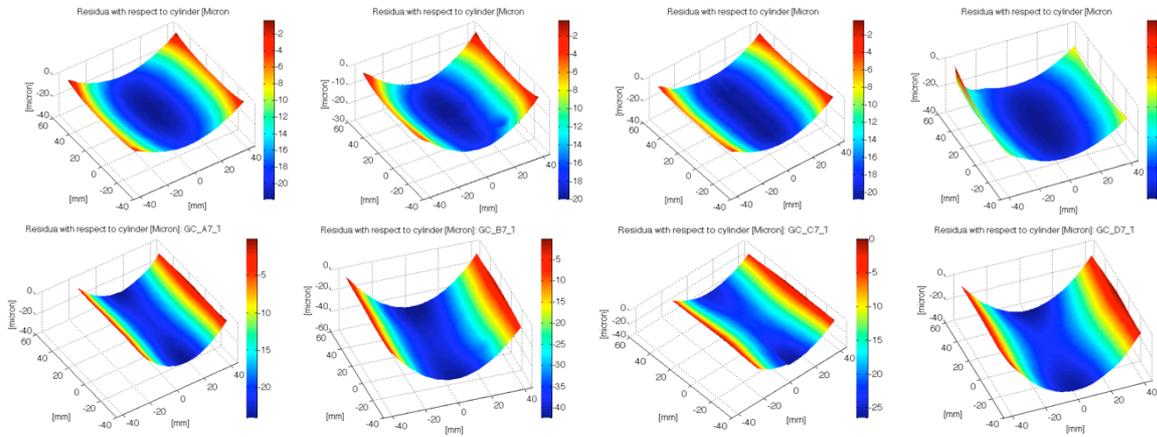

**Fig. 12** Deviation of the measured surface from the nominal cylinder of the slumped Gorilla® glasses, as measured with the CUP: (top) the foils before tempering, (bottom) tempered glasses. From left to right: samples A, B, C and D.

The characteristic parameter values (Peak-to-Valley: PV, rms, for the entire map and along the sole longitudinal direction) of these measurements are summarized in Tab. 3 for each glass foil.

**Table 3** PV and rms, calculated on residual errors after the subtraction of a cylinder with radius of curvature of 1m, before and after chemical tempering. Both values on the entire map and on the longitudinal axis are listed.

|        | BEFORE |      |      |      | AFTER |      |      |      |
|--------|--------|------|------|------|-------|------|------|------|
|        | All    |      | Longitudinal | | All | | Longitudinal | |
| sample | PV     | RMS  | PV   | RMS  | PV    | RMS  | PV   | RMS  |
|        | [µm]   | [µm] | [µm] | [µm] | [µm]  | [µm] | [µm] | [µm] |
| A      | 15.2   | 3.2  | 6.0  | 1.2  | 24.3  | 6.6  | 3.9  | 0.5  |
| B      | 19.6   | 4.5  | 5.7  | 1.1  | 40.1  | 11   | 4.1  | 0.8  |
| C      | 17.6   | 4.1  | 4.8  | 0.8  | 35.4  | 9.5  | 4.6  | 1.1  |
| D      | 20.4   | 4.6  | 6.1  | 1.1  | 35.9  | 9.8  | 5.1  | 1.2  |

Tab. 3 shows high PV values in the entire maps, indicating that the radius of curvature of the four glass foils, after the slumping and the cutting, is rather different from the actual radius of the cylinder of the slumping mould. The main reason is the imperfect CTE matching between the glass and the mould material. Also the pressure setting has still to be optimized in the slumping process for a better mould replication. At the moment, anyway, the different radius of curvature is irrelevant for our purposes, since our integration procedure corrects azimuthal low frequency errors

(Parodi, 2011). Tab. 3 also shows the relevant information in the longitudinal direction, the most critical for our integration procedure. As we already noticed in the 1D measurement (Fig. 10), the chemical tempering reverses the concavity with a reduction of the PV value.

In order to highlight the differences possibly introduced by the tempering process, the measurements before and after tempering were subtracted. The results are shown in Fig. 13 and 14. In Fig. 13, the difference between the maps, obtained before and after tempering is shown, evidencing the saddle introduced by the tempering. In Fig. 14 the difference of the residual maps, obtained before and after tempering after a linear detrend has been applied on both, is shown.

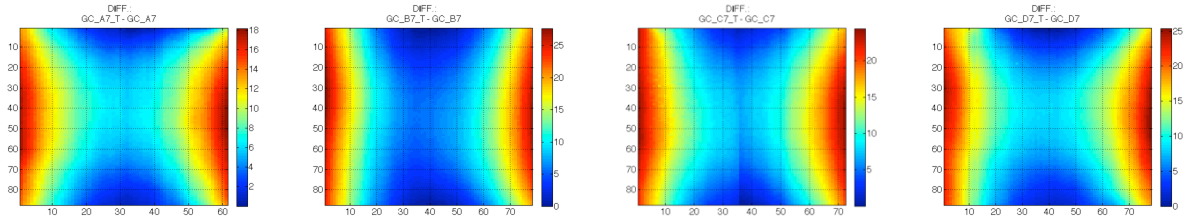

**Fig. 13** Results for the A-B-C-D samples. Difference of the maps shown in Fig. 12, before and after tempering. The z-scale is in µm.

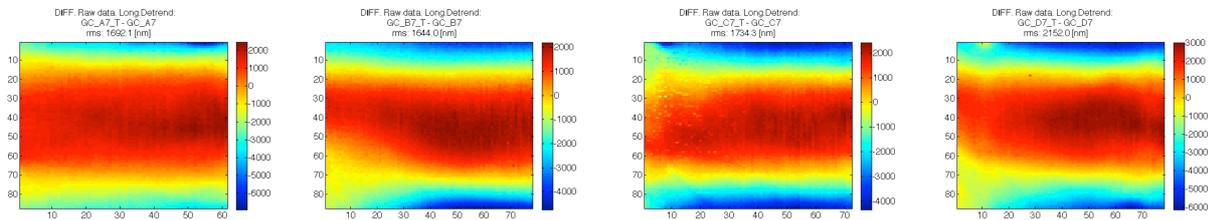

**Fig. 14** Results for the A-B-C-D samples. Difference of the maps shown in Fig. 12, before and after tempering, after a linear de-trend. The z-scale is in nm.

From this analysis we conclude that:
- the results for the four samples are similar in shape and amplitude (Figs. 12-13-14)
- a saddle shape of about 20 µm is introduced during the tempering process (Fig. 13), the major deformation being in the azimuthal direction, which is not critical for our integration
- along the longitudinal scans, a change in concavity by a superimposition of a parabolic profile with an amplitude of about 7-8 µm appears (Fig. 14)
- the overall effect on the longitudinal scans is a amplitude reduction of about 1 µm (Table 3).

In order to evaluate this result in terms of final performances for an X-ray telescope, we have simulated the integration of these glass foils via the modeling described in Sect. 2. Four equally-spaced ribs were considered for the $100 \times 100$ mm$^2$ glass foils. In Fig. 15, we report the integrated map of the B sample, as computed from the CUP measurement, taken before and after the tempering process. The resulting HEW is computed for single reflection with incidence angle of 0.7 degree. The value does not significantly change, going from 6.2 arcsec for the un-tempered glass foil to 6.6 arcsec for the tempered one. For the C sample, the quality of the glass before the tempering was better, giving a value of 5.4 arcsec (Fig. 16). Differently from the B sample, the effect of the chemical tempering on the C sample was substantially degrading the HEW, giving a value after tempering of 8.2 arcsec. The expected HEWs after integration for all the samples are reported in Tab. 4.

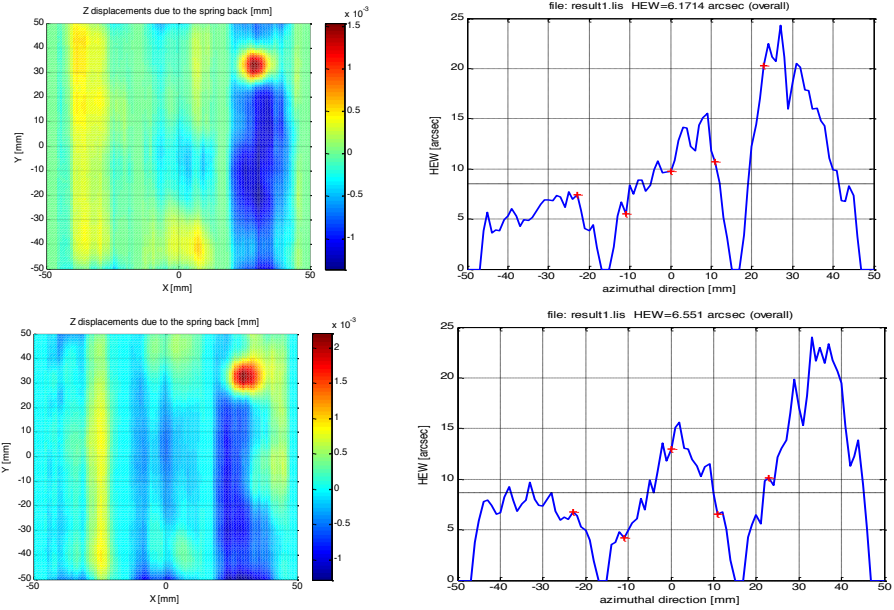

**Fig. 15** Simulation of the integration with 4 ribs for the G7B glass foil, before tempering (top), and after tempering (bottom). Left side: the displacements due to the spring-back of the integrated glass foil. Right side: the expected HEW value, as a function of the x-coordinate, computed for single reflection with incidence angle of 0.7 degree.

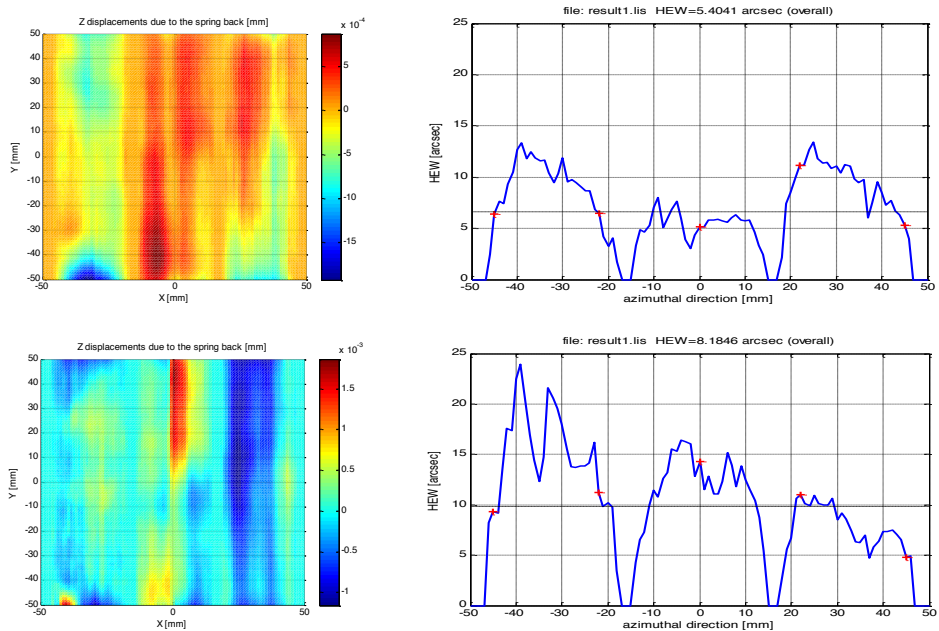

**Fig. 16** Simulation of the integration with 4 ribs for the G7C glass foil, before tempering (top), and after tempering (bottom). Left side: the displacements due to the spring-back of the integrated glass foil. Right side: the expected HEW value, as a function of the x-coordinate, computed for single reflection with incidence angle of 0.7 degree.

| Table 4 HEW expected after integration with 4 ribs | | |
|---|---|---|
| Sample | HEW [arcsec] before tempering | HEW [arcsec] after tempering |
| G7A | 5.5 | 5.8 |
| G7B | 6.2 | 6.6 |
| G7C | 5.4 | 8.2 |
| G7D | 7.7 | 6.6 |
| Mean ± σ | 6.2 ± 1.1 | 6.8 ± 1.0 |

Table 4 shows that the chemical tempering is predicted by simulation to contribute with about 1 arcsec to the degradation of the quality of the X-ray mirrors, computed in single reflection with incidence angle of 0.7 degree. The actual result on the integrated glass foils has still to be proven: for the moment we base the work on the correction capability of our integration process, demonstrated with 200 x 200 mm2 glass foils and 6 ribs (Civitani et al. 2014).

## 10 Conclusion and final remarks

Corning® Gorilla® glasses were selected, for strength reasons, as a viable option for future X-ray telescopes with large effective area and high spatial resolution. They were curved in our laboratories by hot slumping with pressure assistance on a cylindrical mould made of Zerodur K20. Despite the mismatched CTE of the glass and the mould, it was possible to obtain a $200 \times 200$ mm$^2$ slumped glass with profiles suitable for the test. The glass was cut by $CO_2$ laser into four $100 \times 100$ mm$^2$ samples to improve the statistics. They were characterized in shape with 1D and 2D profilometers, both before and after chemical tempering. Most of the difference in the maps, measured before and after tempering, is due to change in the azimuthal direction, which is not critical for our integration, as we efficiently correct in this direction (this was previously proven by X-ray tests on our prototypes, integrated on $200 \times 200$ mm$^2$ size with 6 ribs). In the longitudinal direction, the chemical tempering produces a reduction of the peak to valley of the measured profiles. From the simulation of the integration with 4 ribs, we could compute the expected HEW, at 1keV and single reflection at 0.7 degree incidence angle for on-axis photons on the integrated glass foil. The results of the computation gives only small change of the expected HEW in three cases: G7A, G7B and G7D samples. For sample G7B, for instance, a HEW is expected after integration of 6.2 arcsec before the tempering process, changing to 6.6 arcsec after the tempering. In one case, sample G7C, a higher change in expected HEW after integration is observed, going from 5.4 to 8.2 arcsec, before and after tempering respectively. Finally, the roughness of the tempered Gorilla® glasses was shown to be in line with the roughness of other non-tempered glasses used in our laboratories, and within the prescribed tolerances for the imaging degradation due to scattering.

A possible development of this research would be to quantify the shape deformations after chemical tempering for $200 \times 200$ mm$^2$ cylindrical samples of Gorilla® glass foils, the size as of today used in our laboratories for the production of the X-ray telescopes segment prototypes. The repeatability of the results with several glass foils has to be tested. To improve the slumping result, $Al_2O_3$ material could be used for the slumping mould, owing to its better CTE matching with the Gorilla® glass and to the not-sticking property already proven by our preliminary tests on small flat samples. Destructive tests could be done on slumped Gorilla® samples, comparing the results obtained from tempered and un-tempered samples to quantify the strength improvement after the tempering on our specific slumping process. Finally X-ray tests on the $200 \times 200$ mm$^2$ integrated glass foils should be done to assess the performances of slumped and tempered Gorilla® glass for X-ray telescopes. This work is intended to pave the way for this research.

**Acknowledgments** This research was partially supported by ESA, contract # 22545. We thank Marcos Bavdaz and Eric Wille for the useful discussions. Giancarlo Parodi, Primo Attinà and Enrico Buratti are also acknowledged for very useful discussions. Finally we thank Ronald Stewart from Corning for strength data on Gorilla® glass.